\documentclass[usenatbib]{mn2e}

\usepackage[pdftex]{graphicx}
\usepackage{aas_macros}
\usepackage{times}
\usepackage{amsfonts,amssymb}
\usepackage{thumbpdf}
\usepackage[protrusion,expansion]{microtype}
\usepackage[
    pdftex,
    a4paper,
    plainpages=false,
    pdfpagelabels,
    breaklinks=true,
    bookmarks=true,
    bookmarksopen=true,
    bookmarksopenlevel=1
]{hyperref}

\title{Dark Matter Halo Profiles in Scale-Free Cosmologies}
\author[S.R.~Knollmann, C.~Power and A.~Knebe]
       {Steffen~R.~Knollmann$^1$, Chris~Power$^2$ and Alexander~Knebe$^1$\\
        $^1$Astrophysikalisches Institut Potsdam, An der Sternwarte 16,
            14482 Potsdam, Germany\\
        $^2$Centre for Astrophysics \& Supercomputing, Swinburne
            University of Technology, PO Box 218, Hawthorn,
            Victoria 3122, Australia}

\begin{document}

\newcommand{\dif}{\ensuremath{\mathrm{d}}}
\newcommand{\me}{\ensuremath{\mathrm{e}}}
\newcommand{\mstar}{\ensuremath{M_{*}}}
\newcommand{\AddRef}[1]{\textbf{\it #1}}

\maketitle 

\begin{abstract} 

  We explore the dependence of the central logarithmic slope of dark
  matter halo density profiles $\alpha$ on the spectral index $n$
  of the linear matter power spectrum $P(k)$ using cosmological
  $N$-body simulations of scale-free models (i.e. $P(k) \propto k^n$). These
  simulations are based on a set of clear, reproducible and physically
  motivated criteria that fix the appropriate starting and stopping
  times for runs, and allow one to compare haloes across models with
  different spectral indices and mass resolutions. For each of our
  simulations we identify samples of well resolved haloes in dynamical
  equilibrium and we analyse their mass profiles. By parameterising the 
  mass profile using a ``generalised'' Navarro, Frenk \& White profile
  in which the central logarithmic slope $\alpha$ is allowed to vary 
  while preserving the $r^{-3}$ asymptotic form at large radii, we
  obtain preferred central slopes for haloes in each of our
  models. There is a strong correlation between $\alpha$ and $n$, such 
  that $\alpha$ becomes shallower as $n$ becomes steeper. However, 
  if we normalise our mass profiles by $r_{-2}$, the radius at which
  the logarithmic slope of the density profile is $-2$, we find that 
  these differences are no longer present. This is apparent if we plot
  the maximum slope $\gamma=3(1-\rho/\bar{\rho})$ as a function of 
  $r/r_{-2}$ -- we find that the profiles are similar for haloes
  forming in different $n$ models. This reflects the importance of 
  concentration, and reveals that the concentrations of haloes 
  forming in steep-$n$ cosmologies tend to be smaller than those of
  haloes forming in shallow-$n$ cosmologies. We conclude that there is no
  evidence for convergence to a unique central asymptotic slope, at least
  on the scales that we can resolve.

\end{abstract}
\begin{keywords}

methods: $N$-body simulations -- cosmology: theory -- galaxies: haloes -- 
dark matter

\end{keywords}

\section{Introduction}

The currently favoured model of cosmological structure formation asserts 
that the Universe is dominated by some form of non-baryonic Cold Dark 
Matter (hereafter CDM), and it makes a number of generic predictions about 
how the dark matter clusters on small scales. Arguably the defining 
prediction of the CDM model is that the radial mass density profile of 
dark matter haloes is divergent at small radii, which leads to a central 
density ``cusp''. Characterising the form of this mass density profile has 
been one of the most active research problems in computational 
cosmology over the last decade.

The seminal study of Navarro, Frenk \& White (1996,1997; hereafter NFW) 
introduced the concept of a ``universal'' mass density profile for dark 
matter haloes that form in hierarchical clustering cosmologies. This NFW 
profile is written as

\begin{equation} \label{eq:NFWspherical}
 \rho(r) = \frac{\rho_c}{\left( r/r_s \right)(1+r/r_s)^{2}} \ ,
\end{equation}

\noindent
where $r_s$ is a scale radius and $\rho_c$ is a characteristic density, 
which can be related to $r_s$ once the virial mass of the halo is fixed; 
equation~\ref{eq:NFWspherical} describes a one-parameter family of curves. 
It is convenient to rewrite equation~\ref{eq:NFWspherical} in the more 
general form

\begin{equation} \label{eq:NFWgeneral}
 \rho(r) = \frac{\rho_c}{\left( r/r_s \right)^{\alpha}(1+r/r_s)^{3-\alpha}} \ ,
\end{equation}

\noindent
where $\alpha$ is the central asymptotic logarithmic slope. The NFW 
profile is ``universal'' in the sense that it describes the ensemble 
averaged mass profile of dark matter haloes in dynamical equilibrium, 
independent of virial mass, cosmological parameters and initial power 
spectrum.

During the last decade numerous studies have investigated whether or
not the NFW profile does indeed provide an adequate description of the 
mass profile of dark matter haloes, and to understand the physical
mechanisms that shape the functional form of the profile. NFW recognised that 
$r_s$ correlates with the virial mass of the halo, increasing with 
decreasing virial mass. They argued that this virial mass--scale radius
relation is an imprint of the hierarchical assembly of haloes; low-mass 
haloes tend to collapse before high-mass haloes, when the mean density of 
the Universe is higher, and $r_s$ reflects the mean density of the 
Universe at the time of collapse.

The halo sample that formed the basis of \citet{1997ApJ...490..493N} 
contained of order $10^4$ particles (and hence resolving the profile down 
to about 10\% of the virial radius, according to the convergence criteria 
of \citet{2003MNRAS.338...14P}). Subsequent studies drew upon haloes 
containing about two orders of magnitude more particles within the virial 
radius and confirmed the basic finding of NFW that CDM 
haloes are cuspy (e.g. \citealt{1997ApJ...477L...9F}; 
\citealt{1998ApJ...499L...5M}; \citealt{2000ApJ...529L..69J}; 
\citealt{2000ApJ...535...30J}; \citealt{2001ApJ...557..533F}; 
\citealt{2002ApJ...574..538J}). These studies led to debate about the 
exact value of the logarithmic inner slope, which ranged from $\alpha \sim 
1$~to~$1.5$ (e.g., \citealt{1997ApJ...490..493N}, 
\citealt{1999MNRAS.310.1147M}). The study of
\citet{2003MNRAS.338...14P} provided convergence criteria that allowed 
simulators to understand the impact of numerical artifacts on the central 
structure of haloes, and to identify the innermost radius at which the
mass profile could be considered reliably resolved.
More recently, a great deal of effort has gone into 
even higher resolution simulations that have revealed density profiles of 
CDM haloes to well within $1\%$ of the virial radius 
(\citealt{2004ApJ...606..625F}; \citealt{2004ApJ...607..125T}; 
\citealt{2004MNRAS.349.1039N}; \citealt{2005MNRAS.357...82R}; 
\citealt{2005MNRAS.364..665D}). The highest resolved simulation to-date 
reached an effective mass resolution of about 130 million particles in a 
cluster sized dark matter halo still supporting the evidence for a central 
cusp with a logarithmic inner slope of about $\alpha \sim 1.2$ 
\citep{2005MNRAS.364..665D}.\\

Cosmological simulations allow us to characterise the functional form
of the density profile and to explore the important physical processes
(e.g.  merging, smooth accretion) that drive this form \citep[cf.,
][]{2006MNRAS.368.1931L, 2007ApJ...666..181S}, but a theoretical
understanding of the origin of the mass profile is
essential. Is the form of the profile set by non-linear processes
during the virialisation of the halo, or is there an imprint of the
primordial power spectrum $P(k)$?

One must make strong assumptions about the connection between density
and velocity dispersion to obtain analytical predictions 
about halo structure. For example, under the assumption that the
phase-space density is a power law in radius as suggested by 
\citet{2001ApJ...563..483T} one can 
solve the spherical Jeans equation to obtain the density profile 
(\citealt{2004MNRAS.352L..41H}; \citealt{2005MNRAS.363.1057D}; 
\citealt{2005ApJ...634..756A}). These studies, guided by the results 
of the numerical simulations, confirm the central logarithmic slope to 
be in the range $\alpha \sim 1$ to $2$, although 
\citet{2006JCAP...05..014H} have claimed that equilibrated haloes have 
central slopes of $\alpha \sim 0.8$.

However, it is interesting to ask whether or not there is a dependence
of halo structure on the primordial power spectrum $P(k)$. In
particular, does the form of the profile depend on the (effective)
spectral index $n$ of $P(k)$? One of the key predictions of the NFW
papers is that the shape of the ``universal'' profile should hold in
any hierarchical cosmology, including scale-free ones. This was
confirmed by subsequent numerical simulations
\citep[e.g.][]{1996MNRAS.281..716C}. However, a number of analytical
studies have claimed that the density profile should depend on
spectral index \citep[e.g., ][]{1985ApJ...297...16H,
  1998MNRAS.293..337S, 2000ApJ...538..528S, 2003MNRAS.340.1199H,
  2007ApJ...666..181S}, and these have been supported by numerical
simulations providing similar evidence \citep[e.g.,
][]{1994ApJ...434..402C, 2001ApJ...554..114E, 2003MNRAS.344.1237R,
  2004astro.ph..3352C, 2005MNRAS.357...82R, 2007ApJ...663L..53R}.

Within this context it is interesting to consider the findings of
\citet{2003MNRAS.344.1237R} and \citet{2004astro.ph..3352C}.
Both studies analysed high-resolution cosmological simulations and each 
argued that dark matter haloes at high redshifts have shallow central 
logarithmic slopes, with $\alpha \approx 0.2-0.4$. In particular,
\citet{2003MNRAS.344.1237R} claimed that the central logarithmic slope
depends explicitly on the effective spectral index $n$, in agreement
with the predictions of \citet{2000ApJ...538..528S}. We note that 
\citet[][]{2004ApJ...612...50C} performed and analysed simulations
comparable to those presented in \citet[][]{2003MNRAS.344.1237R} and
found no evidence for the shallow ``cores'' ($\alpha \sim 0.2$);
rather they found that their data favoured ``cusps'' ($\alpha \approx 1$).
However, \citet{2007ApJ...663L..53R} have recently revisited this topic using
simulations with higher mass and force resolution, and they continue to 
argue in favour of shallow cores, in agreement with their earlier study.\\

\citet{2003MNRAS.344.1237R} and \citet{2007ApJ...663L..53R} sought to
isolate the effect of spectral index $n$ on halo structure by using a
series of simulations of the $\Lambda$CDM model with different
box-sizes and evolved for different numbers of expansion factors, to
mimic a single $n$. In this short paper we revisit this topic and
the claims of \citet{2003MNRAS.344.1237R} and
\citet{2007ApJ...663L..53R} using high resolution cosmological 
$N$-body simulations of scale-free models in
which we vary systematically the spectral index $n$. This ``clean''
approach allows us to study whether the density profiles of dark
matter haloes indeed show a dependence on the spectral index $n$.

In \S~\ref{sec:numerical_simulations} we give a detailed description 
of how we have set up and run our scale-free simulations. There are
some important issues to be considered when deciding on \emph{when to
start} a scale-free simulation and \emph{when to stop} the simulation to
compare halo properties. We encapsulate our findings in a set of well
defined and physically motivated criteria. In \S~\ref{sec:analysis} we
present the results of our analysis of the halo mass profiles, and we
summarise our results in \S~\ref{sec:conclusions}.

\section{Numerical simulations}
\label{sec:numerical_simulations}

As we intend to perform scale-free cosmological simulations, we use a
power law for the initial power spectrum of the simulation
\begin{equation}
\label{eqn:definition_Pk}
P\left( k \right) 
= 
a^2 A P_{\rmn{WN}} \left(
                 \frac{k}{k_{\rmn{Ny}}}
               \right)^n,
\end{equation}
with $A$ being the amplitude scaling factor and 
\[
P_{\rmn{WN}} = \left(\frac{L}{N}\right)^3
\quad\rmn{and}\quad
k_{\rmn{Ny}} = \frac{N\upi}{L}
\]
denoting the power of the white noise of a random distribution of
$N^3$ particles in a box with size $L$ and the Nyquist frequency,
respectively.  The simple scaling with expansion factor\footnote{We
  chose $a=1$ at the beginning of the simulations.} $a$ is valid only
for an Einstein--de~Sitter universe, which is the cosmology of our
choice. The box-size $L$ is completely arbitrary because of the
scale-free nature of the power spectra, and so we set it to
$1$~Mpc. $N$ denotes the number of particles along one dimension. We
run a sequence of simulations with the parallel $N$-body code Gadget2
\citep{2005MNRAS.364.1105S} containing $512^3$ particles that differ
only in the value of the spectral index $n$; we use $n$=-0.50, -1.50,
-2.25, -2.50 and -2.75.

In order to set-up and run the simulations, we still need to specify the 
amplitude scaling factor $A$ for all models and to find a criterion for 
when to stop and compare objects from different runs. These two choices will
be presented and discussed in the following two subsections.

\subsection{Initial Conditions}
\label{sec:starting}

For setting up the initial conditions of the simulations, only the
amplitude scaling factor $A$ in equation~\ref{eqn:definition_Pk} needs
to be determined; all other factors are given by the specifics of the
model under investigation, i.e. $N$ and $n$. The mass fluctuations
$\sigma^2_{\rmn{box}}$ within the computational domain can be written
as follows
\begin{equation}
\sigma^2_{\rmn{box}}
=
\left(
  \frac{1}{2\upi}
\right)^3
\int_{\rmn{box}} \dif \bmath{k} \, 
   P\left(k\right)
\end{equation}
where the integration range in $\bmath{k}$ contains all significant
frequencies reproduced in the box, e.g..
$k_{x,y,z} \in \left[2\upi/L,N\upi/L\right]$.
With the definition of $P(k)$ (cf. eq.~\ref{eqn:definition_Pk}) this can
be evaluated analytically to be
\begin{equation}
\label{eq:amplitude}
A
=
\frac{2 \upi^2 \sigma^2_{\rmn{box}}}{k_{\rmn{Ny}}^{3} P_{\rmn{WN}}}
\frac{
n + 3
}{
1 - \left(
           \frac{\sqrt{2}}{N}
    \right)^{n+3}
}
.
\end{equation}
\noindent
Here we use $\sqrt{2}$ rather than $\sqrt{3}$ in the denominator because 
surface diagonals occur more frequently than volume diagonals. Hence by 
specifying $\sigma_{\rmn{box}}$ we also fix the amplitude scaling factor 
$A$. For our simulations we choose $\sigma_{\rmn{box}}$ to be $0.15$ and 
therefore start all models with the same ``integral power'', independent of 
spectral index $n$.

This choice for the normalisation of the initial power spectrum deviates 
from the usual course of action to fix the power at the Nyquist frequency 
$k_{\rmn{Ny}}$ at the white noise level. Our experience tells us that such 
runs start at too late a stage, especially for $P(k)$ with a rather steep 
$n$, because the mass variance $\sigma_{\rm box}$ can be large and 
possibly non-linear. We note that previous studies have arbitrarily 
lowered the amplitude to $1/4$ of the white noise level for such models to 
circumvent this problem (\citealt{1988MNRAS.235..715E}; 
\citealt{1994MNRAS.271..676L}; \citealt{1997ApJ...490..493N}).

\begin{figure}
  \includegraphics[width=84mm]{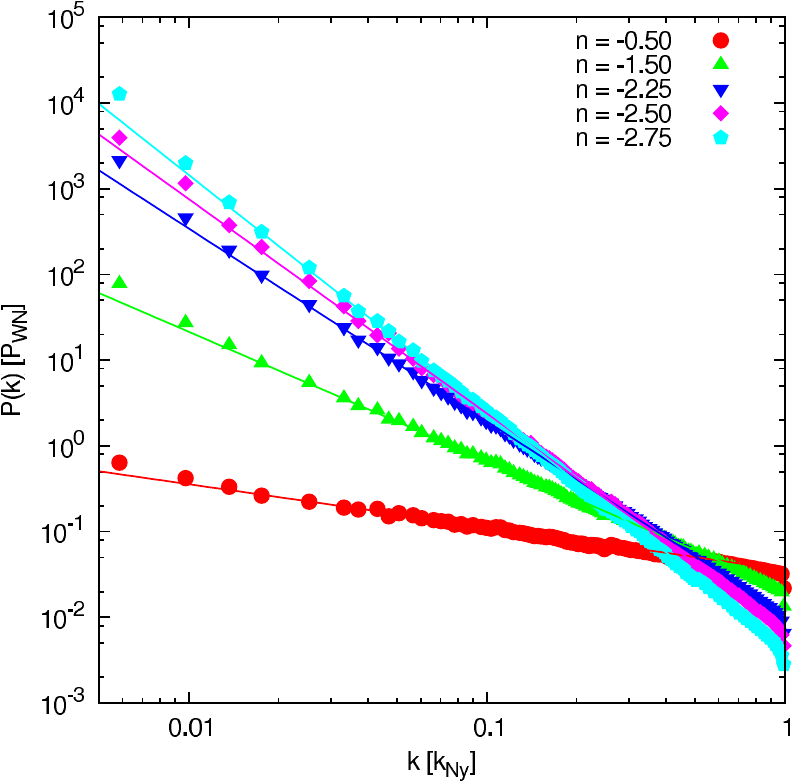}
  \caption{Shown are the theoretical power spectra and their respective
           realisations in the initial conditions of the simulations we
           performed.}
  \label{fig:ICpower}
\end{figure}

Table~\ref{tbl:amplitudes} lists the amplitudes derived from
equation~\ref{eq:amplitude} in terms of the white noise level
$P_{\rmn{WN}}$. Figure~\ref{fig:ICpower} verifies that we recover
the input power spectra from our initial conditions for the selection of
the models under investigation.

\subsection{Stopping Criterion}
\label{sec:stopping_criterion}
In order to define an end-point for our scale-free simulation that
allows for a cross-comparison of identified haloes amongst runs with
different $n$ values we follow \citet{1997ApJ...490..493N} and track
the evolution of $M_{*}(a)$, the typical collapsing mass of an object
in our simulation at expansion factor $a$. This ``non-linear'' mass
scale is defined by requiring that the variance of the linear
overdensity field, smoothed with a top-hat filter enclosing a mass
$M=M_{*}(a)$, should equal the square of the critical density
threshold for spherical collapse (e.g., \citealt{1974ApJ...187..425P},
\citealt{1997ApJ...490..493N})
\begin{equation}
 \sigma^2_{M_{*}} = \delta^2_{\rmn{crit}} .
\end{equation}

The mass variance $\sigma^2_M$ is related to the matter power spectrum 
$P(k)$ via
\begin{equation}\label{sigmarAna}
 \sigma_M^2 = \frac{1}{2\pi^2} \int_0^{+\infty} P(k) \hat{W}^2(k R) k^2 dk ,
\end{equation}
\noindent where $\hat{W}(x)=\frac{3}{x^3} (\sin{x} - x\cos{x})$ is the
Fourier-transform of the top-hat window function, and $R=\left( 3M/4\pi
\rho_{\rmn{crit}}\right)^{1/3}$ relates the spatial scale to a mass scale.

$M_{*}(a)$ is a monotonic function of the cosmic expansion factor $a$
and has a different shape depending on the choice of the spectral
index $n$. We still need to select a fiducial value $M_{*,\rmn{f}}$
(independent of $n$) defining $a_{\rm f}(n)$ via inverting
$M_{*,\rmn{f}}=M_{*}\left(a_{\rm f}(n)\right)$ for each of our scale-free models. A
careful investigation of the maximal allowed expansion factor in the
$n=-2.5$ model leaves us with $M_{*,\rmn{f}} \approx 42000$ particles
(please refer to Appendix~\ref{app:maximalexpansion} for more
details). The adopted $a_{\rm f}(n)$ values are listed in
Table~\ref{tbl:amplitudes}.\footnote{Please note that $a_{\rm f}$ for
  the $n=-2.75$ model is slightly larger than the maximal allowed
  expansion for that model as we used the $n=-2.5$ model as the
  reference for $M_{*,\rmn{f}}$
  (cf. Appendix~\ref{app:maximalexpansion}). We therefore do not
  consider the results derived from this simulation trust-worthy but
  chose to present them anyways.}

\begin{table}
  \caption{The parameters for the initial conditions for the simulations.}
  \label{tbl:amplitudes}
  \begin{tabular}{@{}cccccc}
    \hline
    model & $n$      & $A$ & $a_{\rmn{f}}$ \\
    \hline
    512-0.50 & $-0.50$ & $0.0358$ & $826.0$ \\
    512-1.50 & $-1.50$ & $0.0215$ & $161.5$ \\
    512-2.25 & $-2.25$ & $0.0109$ & $43.9$ \\
    512-2.50 & $-2.50$ & $0.0076$ & $27.3$ \\
    512-2.75 & $-2.75$ & $0.0046$ & $16.7$ \\
    \hline
  \end{tabular}

  \medskip
  $A$ is the amplitude and $n$ the slope of the initial power spectrum.
  The quantity $a_{\rmn{f}}$ gives the adopted expansion of the
  universes that will lead to about 42000 particles in an $M_{*}$ halo. 
\end{table}

\section{Analysis}
\label{sec:analysis}

In this section we first discuss the tool employed to identify haloes 
(section~\ref{sec:halo_identification}) and the selection criteria used
to define the sample of virialised haloes (section~\ref{sec:halo_selection}).
We then employ two approaches to characterise halo structure -- a 
straightforward parameter fit to the density profile to assess the 
dependence of $\alpha$ on $n$ (section~\ref{sec:fitting}), and a
non-parametric estimate of central concentration of the halo based on
$r_{-2}$, the radius at which the logarithmic slope of the density
profile is $-2$ (section~\ref{sec:maximum_slope}). These measures of 
the mass profile allow us to explore the dependence of inner slope on
the spectral index.

\subsection{Halo identification}
\label{sec:halo_identification}

We used an {\small MPI} version of the \texttt{AMIGA} Halo
Finder\footnote{\texttt{AMIGA} is
  freely available from download at
  \url{http://www.aip.de/People/AKnebe/AMIGA/}} (\texttt{AHF}, successor
of \texttt{MHF}
introduced in \citet[][]{2004MNRAS.351..399G}) for identifying haloes
and computing their integral and radial properties. {\small AHF}
locates haloes as peaks in an adaptively smoothed density field using
a hierarchy of grids and a refinement criterion that is comparable to
the force resolution of the simulation (i.e. 5 particles per
cell). Local potential minima are calculated for each of these peaks
and the set of particles that are gravitationally bound to the peaks
are identified as the groups that form our halo catalogue. Each halo
in the catalogue is then processed, producing a range of structural
and kinematic information. Haloes are defined such that the virial
mass is $M_{\rm vir}=4 \pi \rho_{\rm crit} \Delta_{\rm vir} r_{\rm
  vir}^3/3$, where the overdensity criterion $\Delta_{\rm vir}=178$
for an Einstein-de~Sitter cosmology.

\subsection{Halo selection}
\label{sec:halo_selection}

We restrict our analysis to only those haloes that contain in excess of 
$3.15 \times 10^4$ particles($\simeq 0.75 \mstar$) within the virial 
radius. Furthermore we employ a criterion to exclude objects that we do 
not expect to be in dynamical equilibrium. For this we use $\Delta_r$, the 
displacement of the centre of mass of all material within the virial 
radius with respect to the centre of potential of the halo, normalised to 
the virial radius of the halo;

\begin{equation}
\label{eq:deltar}
\Delta_r=\frac{|\vec{r}_{\rm cm}-\vec{r}_{\rm cen}|}{r_{\rm vir}}
\end{equation}

\noindent We reject all haloes for which $\Delta_r > 0.05$. This is a more 
conservative requirement than has been used in other recent studies 
\citep[e.g.][]{2007MNRAS.381.1450N}. However, we show in another paper 
(Power~et~al, in prep.) that this ensures that a halo is unlikely to 
have experienced a merger with mass ratio greater than 10\% in the 
previous dynamical time (defined at the virial radius).

Additionally we define a subset of the catalog, consisting of haloes at 
about \mstar: we use objects in the mass range of $0.75\mstar$ to 
$1.5\mstar$ which corresponds to halos with a number of particles $N\in 
[3.15,6.30]\times 10^4$. In table~\ref{tbl:halo_select} we give a summary 
of the number of haloes used in our analysis.

\begin{table}
  \caption{Definition of various halo samples.}
  \label{tbl:halo_select}
  \begin{tabular}{@{}lccc}
    \hline
    Run     & all
	    & $M_{*}$ sample 
	    & high-mass sample \\
    \hline
    512-0.50 & 276 & 186 & 38 \\
    512-1.50 & 172 & 107 & 44 \\
    512-2.25 & 119 &  62 & 33 \\
    512-2.50 &  64 &  35 & 19 \\
    512-2.75 &  38 &  15 & 16 \\
    \hline
  \end{tabular}

  \medskip

  The columns give the number of haloes that satisfy our selection 
  criteria. The first column gives the total number of objects consisting 
  of more than $3.15\times 10^4$ particles; we refer to it as ``all''. The 
  second column gives the number of objects in our $M_{*}$ sample, i.e. 
  haloes with masses in the range $M\in [0.75, 1.5] M_{*}$. Finally we 
  list the number of haloes with more than $N>10^{5}$ particles, which we 
  refer to as ``high-mass sample''. Note that all haloes satisfy our the 
  centre-of-mass offset criterion and are therefore considered to be in 
  dynamical equilibrium.

\end{table}

\subsection{The Preferred Slope \texorpdfstring{$\alpha$}{alpha}}
\label{sec:fitting}

We begin our study of the dependence of the central logarithmic slope 
$\alpha$ on the spectral index $n$ by fitting the generalised NFW
profile (cf. equation~\ref{eq:NFWgeneral}) to each halo in our $M_{*}$ sample.  Recall that 
\begin{equation} 
\label{eq:extend_nfw}
\rho\left(r\right) =
\frac{\rho_c}
     {\left(\frac{r}{r_s}\right)^\alpha
      \left(1+\frac{r}{r_s}\right)^{3-\alpha}}
\end{equation}

\noindent where $r_s$ and $\rho_c$ are the fitted parameters. We
perform several fits for each profile, varying $\alpha$ in the range of
$\alpha = 0.3,\dots,1.8$. The case of $\alpha = 1.0$ corresponds to the
NFW profile.

For each halo we determine a set of parameters for each choice of
$\alpha$. By comparing the $\chi^2$-value we can, in principle, find
the preferred $\alpha$ for a halo. However, we are interested in 
whether or not $\alpha$ depends on $n$ for a \emph{typical} halo, and 
so a statistical analysis is required. Therefore, we define a quality 
measure of the fit,
\begin{equation}
Q_j = \frac{\sum_{\mathrm{bins}} 
      \left( \frac{X_{ij} - \rho_\mathrm{fit}\left(r_i\right)}
                  {X_{ij}} 
       \right)^2}{\mathrm{dof}};
\end{equation}
\noindent 
here $\mathrm{dof}$ denotes the number of degrees of freedom of the
fit, $X_{ij}$ the density value of the $j$th halo in the $i$th radial bin and
$\rho_\mathrm{fit}$ the corresponding value of the analytic function at
radius $r_i$. 

\begin{figure}
  \includegraphics[width=84mm]{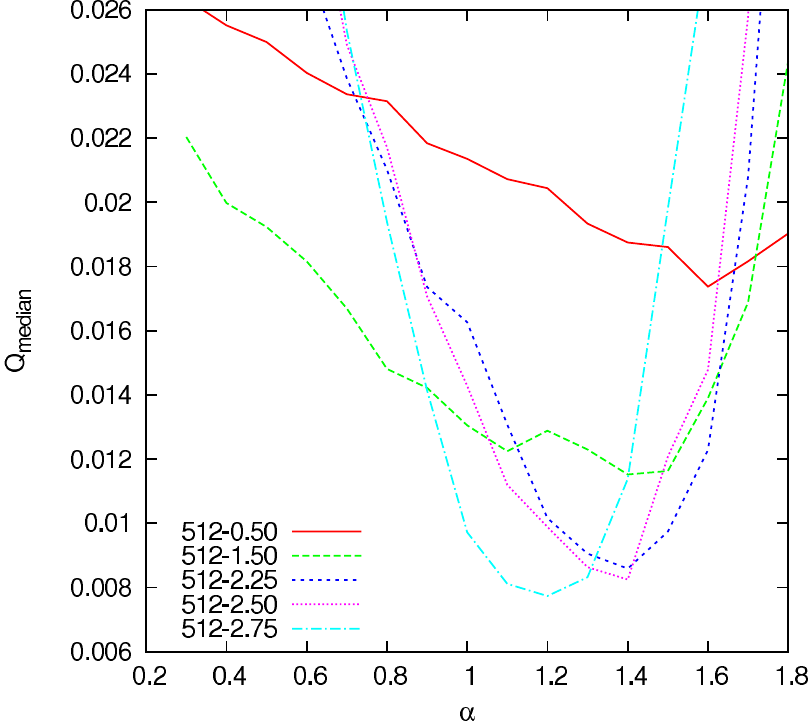}
  \caption{The median of the quality in the different simulation boxes
           for objects in the $M_{*}$ sample.}
  \label{fig:qvsalpha}
\end{figure}

Taking the median $Q$ in each simulation box for each value of $\alpha$ we
can explore how the preferred inner slope of the density profile
depends on the spectral index $n$, as shown in
Figure~\ref{fig:qvsalpha}. Each curve indicates how the quality of the
fit varies with $\alpha$ for a given
$n$, and the minima of the curves pick out the preferred $\alpha$ for
a particular $n$. For the $n=-0.5$ run we find a minimum at  $\alpha
\simeq 1.6$, while for the $n=-2.75$, the preferred $\alpha \simeq 1.2$.
Even though the curves for the $n=-1.5$ ($-2.25$, $-2.5$) runs seem to
have their minima at $\alpha \simeq 1.4$, a trend can be seen.
Therefore we might conclude that the preferred $\alpha$ decreases
with decreasing $n$; that is, the steeper the spectral index, the
shallower the central logarithmic slope.

However, care must be taken when interpreting slopes. Close inspection
of equation~\ref{eq:extend_nfw} will reveal that the logarithmic slope
will depend on the scale radius $r_s$. Therefore, an apparent trend in
$\alpha$ with $n$ may instead be attributable to a trend in $r_s$ with
$n$. If $r_s$ is systematically smaller in a model with spectral index
$n_1$ when compared to a model with spectral index $n_2$, haloes will
be systematically more concentrated and so fits to the resolved part of
the density profile will produce higher ``effective'' slopes $\alpha$. 
We investigate this further in \S~\ref{sec:maximum_slope}.

\subsection{The Maximum Slope \texorpdfstring{$\gamma$}{gamma}}
\label{sec:maximum_slope}

A useful non-parametric measure of halo structure is the maximum
asymptotic slope,
\begin{equation}\label{eq:maxslope}
\gamma\left(r\right) = 3\left( 1 - \frac{\rho(r)}{\bar{\rho}(r)} \right),
\end{equation}
\noindent which uses the local mass density $\rho$ and the enclosed
mass density $\bar{\rho}$ to derive an upper limit to the logarithmic slope
of the density profile at radius $r$. Equation~\ref{eq:maxslope}
assumes that the halo is spherically symmetric with a density profile
interior to $r$ given by a power-law, $\rho(r) \propto
r^{-\gamma}$. $\gamma$ defines an upper limit to the slope; a steeper
slope would require more mass interior to $r$ than is measured. 
We note that this measure was used by \citet{2004MNRAS.349.1039N} for
resimulated haloes of different masses but comparable particle
resolution in their study of the universality of the mass profile.

\begin{figure}
  \includegraphics[width=84mm]{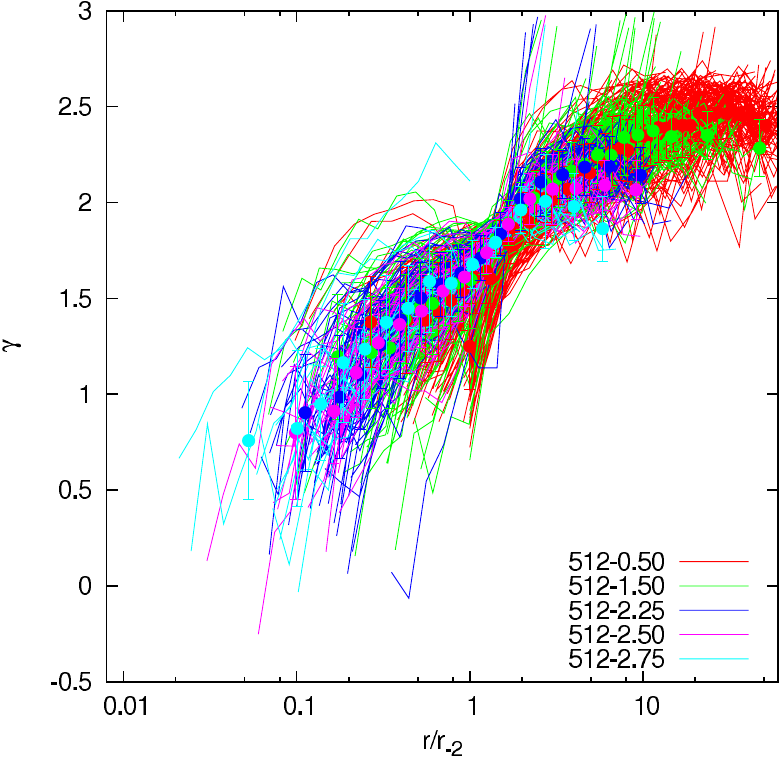}\\
  \includegraphics[width=84mm]{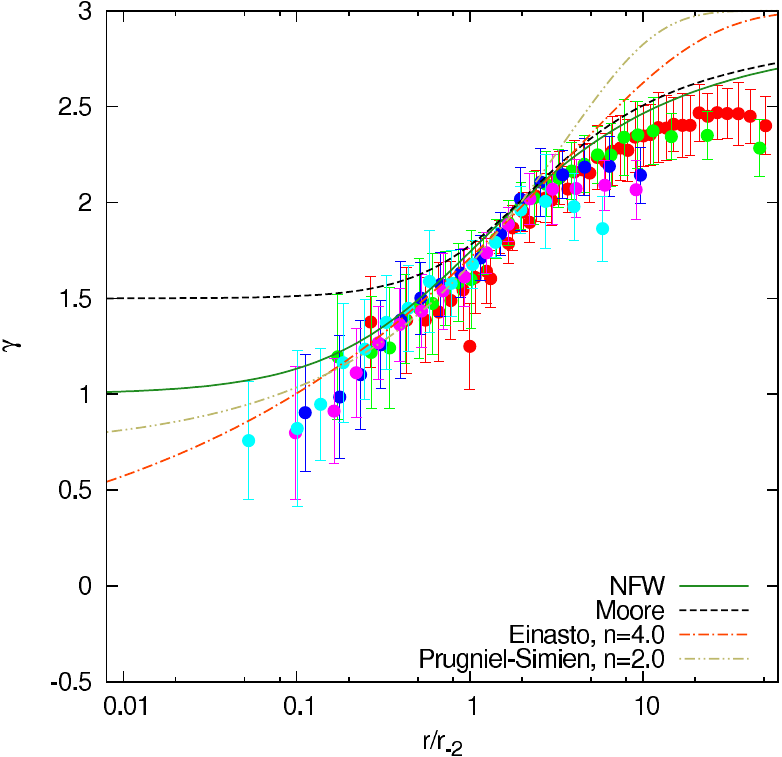}
  \caption{The maximal inner slope as given by equation~\ref{eq:maxslope}
           as a function of the radius normalised to $r_{-2}$ for 
           haloes in the $M_{*}$ sample.
           Upper panel: We plot all curves from all the simulations and
           over-plot a binned curve for each run.
           Lower panel: Here only the averages of each simulation are
           shown and compared to a set of theoretical predictions
           derived from the NFW, Moore et al., Einasto and 
           Prugniel-Simien profiles. Note that in this instance $n$
           corresponds to the shape parameter for the Einasto and 
           Prugniel-Simien profiles.}
  \label{fig:maxslope_rr2}
\end{figure}

In Figure~\ref{fig:maxslope_rr2} we plot the radial variation of
$\gamma$ for all haloes that satisfy the selection criteria of 
\S~\ref{sec:halo_selection} and are in the same mass-range as those used
in section~\ref{sec:fitting}, colour-coded according to the spectral
index $n$ of the model in which they form. Note that we have normalised
these profiles to $r_{-2}$, the radius at which the differential
mass profile $\rho\,r^2$ reaches a maximum. For a NFW profile, $r_{-2}$
is identical to the scale radius $r_s$ and so it provides an attractive
non-parametric measure of concentration\footnote{We have taken great
  care to ensure that our estimates of $r_{-2}$ are reliable. As haloes 
  become more concentrated, $r_{-2}$ can rapidly approach the
  innermost reliably resolved radius, which can affect the accuracy
  with which $r_{-2}$ is estimated. This issue is particularly acute
  for haloes in the shallow-$n$ cosmologies. Therefore we do not
  consider haloes less massive than $0.75 M_{*}$.}.

When normalising the radius in this manner, we find excellent agreement 
between the average shapes of the $\gamma$ profiles between the
different $n$ models. However, the scatter between profiles within a
given simulation is significant (cf. upper panel of
figure~\ref{fig:maxslope_rr2}), which strengthens our argument that it
is essential to use a statistical sample of haloes when discussing the
asymptotic inner slope. It is noticeable that the average profile in
each model we have looked at continues to becomes shallower with
decreasing radius, without showing evidence for convergence to an
asymptotic value \citep[c.f.][]{2004MNRAS.349.1039N}. We find similar 
behaviour when considering only haloes in the high-mass sample.\\

This figure also confirms our suspicion that it is the scale radius
$r_{-2}$ or concentration $c_{\rm vir}=r_{\rm vir}/r_{-2}$ rather than
the slope $\alpha$ that varies with $n$. We find that haloes forming
in the $n=-0.5$ model tend to be more concentrated
(cf. figure~\ref{fig:concentration} below) than haloes forming in runs with
steeper spectral indices. Therefore fits with a generalised NFW
profile (cf. equation \ref{eq:extend_nfw}) tend to favour smaller
values of $\alpha$ for steeper $n$ because these haloes tend to be
less concentrated, and so we resolve the profile to smaller fractions
of $r_{-2}$, where the flattening of the profile is more
apparent. Therefore a shallower effective slope $\alpha$ will tend to be
preferred.

It is a relatively straightforward exercise to obtain expressions for
equation~\ref{eq:maxslope} for the NFW profile and the
\citet{1998ApJ...499L...5M} profile. Two other analytical model
profiles have been promisingly applied to halo density profiles, the
\citet{1965Einasto} and \cite{1997A&A...321..111P} profiles, which
provide better fits than the NFW profile. \citet{2006AJ....132.2685M}
argue that the Einasto model performed best in fitting halo profiles,
followed closely by the Prugniel-Simien model. Expressions for
$\gamma$ for the Einasto and Prugniel-Simien models are given in the
Appendix~\ref{app:maximumslope}.

In the lower panel of figure~\ref{fig:maxslope_rr2} we over-plot the
averaged $\gamma$-curves with the theoretical predictions derived for
the four analytic profiles mentioned above. We find that the
Moore profile is unable to reproduce the observed behaviour. The NFW 
profile is consistent with our data for the 512-0.50 and 512-1.50
runs, but it fails to reproduce the continual flattening of $\gamma$ to
small radii. In contrast, both the Einasto and Prugniel-Simien profiles
capture the behaviour of our data well at small radii\footnote{Note that in
this instance $n$ corresponds to the shape parameter for the Einasto and 
Prugniel-Simien profiles.}. Interestingly we note that all of the
analytical profiles tend to overestimate the slope of the density
profile at large $r/r_{-2}$, which appears to roll over
and flatten off. This is most apparent for the data points from the
$n=-0.5$ run.\\

\begin{figure}
  \includegraphics[width=84mm]{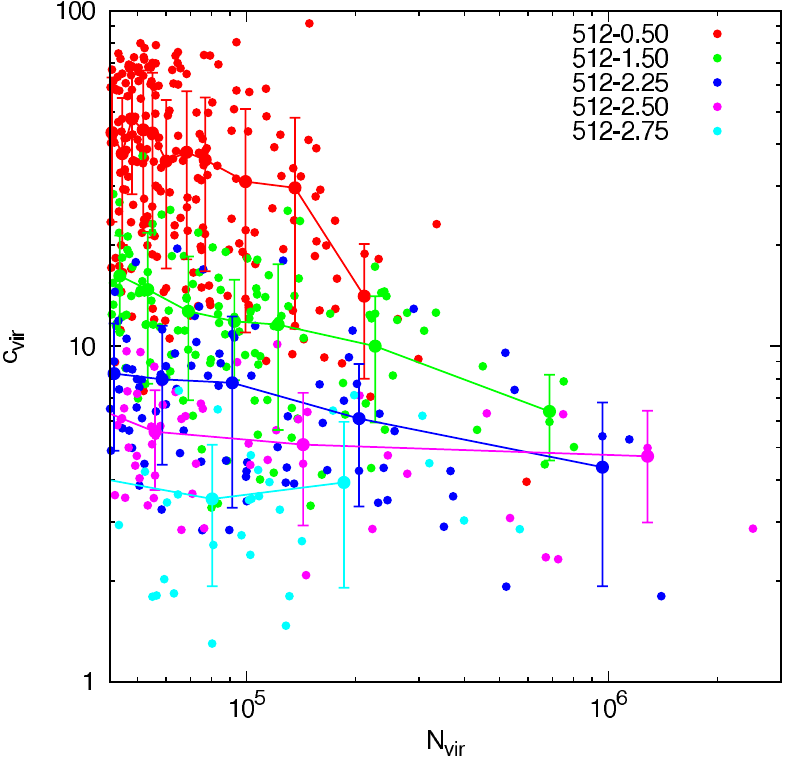}\\
  \includegraphics[width=84mm]{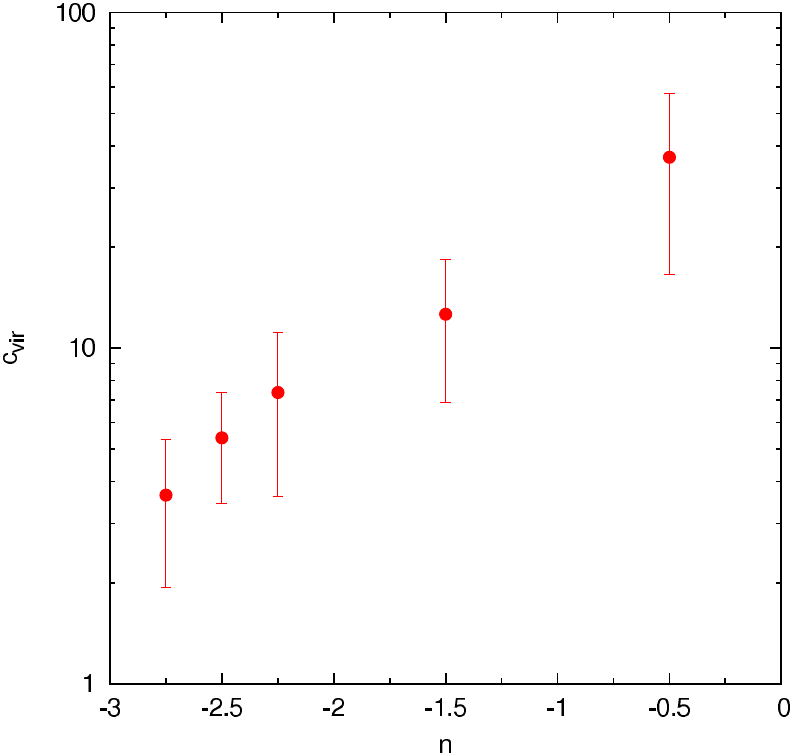}
  \caption{In the upper panel we show the concentration of all haloes
           with mass larger than \mstar\ as a function of the number of
           particles within the virial radius. Additionally we plot a
           binned distribution of the concentration. The lower panel
           gives the mean concentration for objects in our $M_{*}$ sample depending on the
           spectral index $n$.}
  \label{fig:concentration}
\end{figure}

In Figure~\ref{fig:concentration} we make explicit the connection
between $r_{-2}$ and the concentration $c_{\rm vir}=r_{\rm
  vir}/r_{-2}$, showing how $c_{\rm vir}$ varies with halo mass (given by 
the number of particles within the virial radius $N_{\rm vir}$; upper 
panel), and the spectral index $n$ (lower panel). A similar figure can be 
found in~\cite{1997ApJ...490..493N}, who looked at scale-free models with 
spectral indices of $n=-0.5, -1.0, -1.5$, but who derived their 
concentrations from fits of NFW profiles. Although our concentrations are 
calculated in a non-parametric manner, it is reassuring that we see a 
similar trend to that reported in ~\cite{1997ApJ...490..493N}. The 
variation of concentration with mass (upper panel) suggests that this 
relation may be steeper in cosmologies with shallower spectral indices, 
which in turn may reflect the importance of merging for the growth of the 
most massive haloes in these cosmologies. Although the trends are 
tentative, we believe that this is an extremely interesting figure and one 
that may allow to better understand the physical origin of the density 
profile. We shall return to this issue, albeit briefly, in 
\S~\ref{sec:conclusions}.

The trend for concentration to decrease with steeper spectral index is 
more robust, and we show this explicitly by plotting the mean 
concentration of haloes in the $M_{*}$ sample versus spectral index (lower 
panel). This makes clear our assertion that it is the concentration that 
depends strongly on the spectral index.

\section{Summary \& Conclusions}
\label{sec:conclusions}

The primary motivation for this paper has been to establish what
dependence, if any, the logarithmic slope $\alpha$ of the inner mass density
profile of dark matter haloes has on the spectral index $n$ of the
linear matter power spectrum. For this purpose we have run a series of 
high resolution cosmological $N$-body simulations of scale-free models
(i.e. $P(k) \sim Ak^n$) with values of the spectral index $n$ varying
between $-0.5 \geqslant n \geqslant -2.75$. By using scale-free models
and fixing $n$ in this manner, the problem of identifying correlations 
between $\alpha$ and $n$ becomes a relatively straightforward one. 
Relatively, because there is some freedom in the choice of criteria one 
can use to determine when to start and when to finish scale-free
simulations. 

To address this, we have derived clear, reproducible and physically 
motivated criterion that allows us to set up scale-free simulations and 
that is presented in \S~\ref{sec:starting}.
In short, we start the simulations 
with the same integral power in the $k$-range defined by the the total 
number of particles $N$; it is especially important to have an objective 
criterion for starting simulations when studying mass profiles. To
determine when to stop our simulations, we follow
\citet{1997ApJ...490..493N} and use the evolution of the typical
collapsing mass $M_{*}$. We do so once the mass in a typical $M_{*}$
halo corresponds to approximately $42000$ particles.

Having established a set of well-defined criteria to set up and run 
cosmological simulations of scale-free models, we performed a sequence of 
high resolution runs ($1 h^{-1} \rm Mpc$ boxes, $512^3$ particles) that
we used to investigate the dependence of the central logarithmic slope
of the dark matter halo density profile $\alpha$ on the spectral index $n$.
We varied $n$ between $-0.5$ and $-2.75$ and selected samples of well
resolved haloes ($N_{\rm vir} \geqslant 3.15 \times 10^4$) in dynamical
equilibrium in each run (ranging from $\sim 270$ haloes in the $n=-0.5$
run to $\sim 40$ haloes in the $n=-2.75$ run). Using $\chi^2$ fits to a
generalised NFW profile, we identified preferred values of the inner
slope $\alpha$ and indeed, we found a trend for the inner slope to become
shallower with steeper spectral index -- from $\alpha \simeq 1.6$ for
$n=-0.5$ to $\alpha \simeq 1.2$ for $n=-2.75$.

However, we argue that it is not the central slope $\alpha$ that depends 
on spectral index; rather, it is the scale radius of the halo, $r_s$, or 
our preferred measure $r_{-2}$, the radius at which the differential mass 
profile $\rho\,r^2$ reaches its maximum value. We have shown that haloes 
in different $n$ models have similar radial profiles of the maximum slope 
$\gamma=3(1-\rho/\bar{\rho})$ when normalised to $r_{-2}$. However, as 
already shown by~\citet{1997ApJ...490..493N}, haloes that form in models 
with steeper $n$ tend to be less centrally concentrated than haloes 
forming in models with shallower $n$, and so their mass profiles can be 
resolved to smaller fractions of $r_{-2}$, where the flattening of the 
profile is more apparent. Haloes in these models would then appear to have 
shallower central profiles.\\

We noted in the introduction that using scale-free simulations to study 
the effect of the spectral index of the power spectrum on the central 
structure of dark matter haloes was preferable to the approach taken by 
\citet{2003MNRAS.344.1237R}, \citet{2004astro.ph..3352C}, 
\citet{2004ApJ...612...50C}, and \citet{2007ApJ...663L..53R}, in which the 
box size and analysis redshift were varied to capture the behaviour of 
different spectral indices. The claims of \citet{2007ApJ...663L..53R} are 
of particular interest; these authors argue that halo concentration is a 
universal constant and that dwarf galaxies identified at $z$=10 have 
logarithmic slopes shallower than $0.5$.

Our results strongly disagree with these claims. Although the central 
slope $\alpha$ that we obtain by fitting a generalised NFW profile does 
vary with $n$, this does not imply that the shape of the profile is 
sensitive to $n$, for the reasons presented above. Moreover, as we show in 
Figure \ref{fig:maxslope_rr2}, density profiles normalised by $r_{-2}$, 
which is equivalent to concentration, have similar shapes on average, 
independent of spectral index; this implies that it is $r_{-2}$ and 
consequently concentration that depends on $n$. It is possible that our 
profiles for $\gamma(r)$ may roll over and approach different asymptotic 
value at smaller radii than we can resolve with our simulations, but over 
the range of radii that we can reliably resolve we find no evidence for 
such behaviour. It should be noted that we checked for this by producing 
figure~\ref{fig:maxslope_rr2} for haloes in the high-mass sample (cf. 
Table~\ref{tbl:halo_select}) but we did not find any evidence for 
convergence.\\

We briefly consider the dependence of $r_{-2}$ on $n$. We show in 
figure~\ref{fig:concentration} that haloes forming in cosmologies with 
steeper spectral indices tend to be less centrally concentrated than 
haloes forming in cosmologies with shallower spectral indices, confirming 
and extending results presented by~\citet{1997ApJ...490..493N}. This is 
interesting because we expect that haloes forming in shallow-$n$ models do 
so predominantly by accretion, whereas haloes in steep-$n$ models form via 
merging.  If we examine the mass concentration relation in shallow-$n$ 
models (upper panel in figure~\ref{fig:concentration}), we find that more 
massive haloes tend to be less centrally concentrated than their less 
massive counterparts. In contrast, we find that this relation is far less 
pronounced in steep-$n$ models.

Prescriptions for halo concentration, such as \citet{2001ApJ...554..114E}, 
assert that concentration is determined by the halo's collapse redshift -- 
haloes that collapse earlier do so when the mean density of the Universe 
is higher, and so they tend to be more centrally concentrated. It has been 
argued that this reflects the rate at which the proto-halo accretes mass
\citep[cf.][]{2003ApJ...597L...9Z}. 
Haloes are said to assemble their mass in two distinct phases -- 
fast-accretion phase during which the depth of the potential well is set, 
and a slow-accretion phase during which mass is added to the outer parts 
of the halo \citep[e.g][]{2006MNRAS.368.1931L}. This has important 
implications for the concentration. For example, 
\citet{2002ApJ...568...52W} argued that during the fast accretion phase 
the central density is set by the background density, but as the accretion 
rate slows and the halo enters its slow accretion phase, the central 
density stays approximately constant and the halo's concentration grows in 
step with the virial radius.  Merging also plays an important role; for 
example, \citet{2003ApJ...593...26M} looked at the effect of major mergers 
on concentration and found that violent relaxation drives the remnant's 
density profile towards a form that is essentially identical to the one it 
would have formed through pure accretion.

What can we learn about the physical origin of the density profile from 
our simulations? As we noted above, haloes in shallow-$n$ models grow 
predominantly by accretion, whereas those in steep-$n$ models grow via 
merging. This link between spectral index and accretion rate can be made 
explicit \citep[see, for example,][]{2006MNRAS.368.1931L}, which makes the 
scale-free simulation a very interesting tool with which to study the 
origin of the profile. The results -- that concentrations tend to be 
higher in shallow-$n$ models, and that concentration decreases with 
increasing virial mass in these models (assuming that merging plays a role 
in the growth of the most massive halos) -- indicate that ``two-phase 
accretion'' may be very important. However, to properly address this 
question a more detailed study is required. We are currently undertaking 
such a study, an account of which will be presented in a future paper.

\section*{Acknowledgements} 
\label{sec:acknowledgements} 

SRK and CP thank Greg Poole for useful discussions over coffees at Cafe FM 
on Glenferrie Road in Melbourne. SRK acknowledges financial support from 
the Centre for Astrophysics and Supercomputing's visitor programme during 
the writing of this paper. SRK and AK acknowledge funding through the Emmy 
Noether Programme by the DFG (KN 755/1). CP acknowledges funding through 
the Australian Research Council funded ``Commonwealth Cosmology 
Initiative'', DP Grant No. 0665574. All of the simulations and analyses 
were carried out on the Sanssouci and Luise clusters at the AIP.

\vspace{1cm} \bsp

\bibliographystyle{mn2e}
\bibliography{paper}

\appendix

\section{Maximal Expansion}
\label{app:maximalexpansion}

A natural stopping criterion for a scale-free cosmological simulation
is given by the requirement that the fundamental mode (or a multiple
of it) turns non-linear; this defines a maximal expansion
$a_{\rmn{max}}$ \citep[cf. ][]{1988MNRAS.235..715E}. But we like to
remind the reader that in this study we compared simulations run with
varying $n$ values and hence had to define a different stopping
criterion that allows us to cross-compare those runs at a similar
evolutionary stage.

However, in order to determine the maximal allowed expansion to obtain
the fiducal $M_{*, \rmn{f}}$ value we follow
\citet{1988MNRAS.235..715E} and define a critical wavenumber $k_0$,
corresponding to a filtering scale for which the rms density
fluctuations approach unity
\begin{equation} \label{eq:rms}
\left<
       \sigma^2
\right>_{k_0} 
=
\left(
      \frac{1}{2\upi}
\right)^3
\int_{\mathbb{R}^3} \dif \bmath{k} \,
P\left(k\right)
\me^{-\left(
            k/k_0
      \right)^2}
=
1 .
\end{equation}
\noindent
This equation divides the power spectrum in two parts: for all
frequencies below the critical frequency $k_0$ the growth is still linear
and for all higher frequencies the growth is non-linear. Solving
equation~\ref{eq:rms} for $k_0$ yields:
\begin{equation}
k_0 =  k_{\rmn{Ny}}
       \left[
          \frac{\upi}{4} A a^2 \Gamma\left(\frac{n+3}{2}\right)
       \right]^{-\frac{1}{n+3}}
.
\end{equation}
\noindent
Therefore the evolution of the critical frequency $k_0$ with the
expansion factor $a$ of the universe is now known and can be inverted
to calculate a maximal $a_{\rmn{max}}$ as a function of a pre-defined
threshold wavenumber $k_0$
\begin{equation}
\label{eq:amax}
a_{\rmn{max}}
=
\left[
      \frac{\upi}{4} A x^{n+3} \Gamma\left(\frac{n+3}{2}\right)
\right]^{-\frac{1}{2}}
\end{equation}
where $x$ is given by the relation
\begin{equation}
 k_0 = x k_{\rmn{Ny}}
.
\end{equation}

As the threshold frequency we choose the smallest wavenumber that we
are able to resolve numerically, i.e. the fundamental mode in our
computational domain as given by $x = 2/N$. The maximal allowed
expansion factors for our scale-free models are listed in
Table~\ref{app:amax} and we encourage the reader to compare them
against the adopted expansion factors summarized in
Table~\ref{tbl:amplitudes}. Our stopping criterion complies with the
requirement that the fundamental mode is still in the regime of linear growth.

\begin{table}
  \caption{The maximal allowed expansion factors.}
  \label{app:amax}
  \begin{tabular}{@{}cccccccc}
    \hline
    model & $n$      & $a_{\rmn{max}}$ \\
    \hline
    512-0.50 & $-0.50$ & $6413.5$ \\
    512-1.50 & $-1.50$ & $445.0$  \\
    512-2.25 & $-2.25$ & $56.2$   \\
    512-2.50 & $-2.50$ & $27.3$   \\
    512-2.75 & $-2.75$ & $12.1$   \\
    \hline
  \end{tabular}

\end{table}

\section{Maximum Slope : Expressions for Einasto and Prugniel-Simien Models}
\label{app:maximumslope}

For the Einasto model, the density profile can be written as
\begin{equation}
\hat{\rho}_\mathrm{Ein}(\hat{r}) = \exp\left\{
       -2n\left(\hat{r}^{\frac{1}{n}} - 1\right)
                          \right\}
\end{equation}
\noindent where $\hat{\rho}$ and $\hat{r}$ are the density and the
radius in units of $\rho_{-2}$ and $r_{-2}$, respectively, and $n$
gives the shape of the density profile. We will use
this notation throughout this appendix. For formulations of the profiles
in various units and conversions between them see
\citet{2006AJ....132.2685M} and \citet{2006AJ....132.2701G}.

From this, the enclosed mass $\hat{M}_\mathrm{Ein}$
can be obtained. This is given by:
\begin{equation}
  \hat{M}_\mathrm{Ein} = 2\pi\me^{2n} \left(2n\right)^{1-3n}
  \gamma\left(3n,2n \hat{r}^\frac{1}{n}\right)
\end{equation}
\noindent
where $\gamma(a,x)$ is the lower incomplete Gamma Function given by
\begin{equation}
\gamma(a,x) = \int_0^x t^{a-1} \me^{-t}\,\dif t \,.
\end{equation}
This yields
\begin{equation}
  \gamma\left(\hat{r}\right) =
  3\left( 1-\frac{2}{3}
  \frac{(2n)^{3n-1} \hat{r}^3 \me^{-2n\hat{r}^{\frac{1}{n}}}}
       {\gamma\left(3n,2n \hat{r}^\frac{1}{n}\right)}
       \right)
\end{equation}
\noindent
for the maximum slope. 

For the Prugniel-Simien model, the density profile is written as 
\begin{equation}
\hat{\rho}(\hat{r}) = \hat{r}^{-p}
                      \exp\left\{
       -n(2-p)\left(\hat{r}^{\frac{1}{n}} - 1\right)
                      \right\}
\end{equation}
\noindent
where $p$ is a function of $n$ which can be approximated by
\begin{equation}
p(n) = 1 - 0.6097/n + 0.05463/n^2 \, .
\end{equation}
The enclosed mass profile can be written as 
\begin{equation}
  \hat{M}\left(\hat{r}\right) =
  \frac{4\pi}{2-p}\me^{n(2-p)}\left(n(2-p)\right)^{1-n(3-p)}
  \gamma\left(n(3-p),(2-p)n\hat{r}^{\frac{1}{n}}\right)
\end{equation}
\noindent which leads to 
\begin{equation}
  \gamma\left(\hat{r}\right) =
  3\left( 1-\frac{2-p}{3}
  \frac{\left(n(2-p)\right)^{n(3-p)-1}
    \me^{-n(2-p)\hat{r}^{\frac{1}{n}}}}
       {\hat{r}^{p-3} \gamma\left(n(3-p),(2-p)n\hat{r}^{\frac{1}{n}}\right)}
       \right)
\end{equation}
for the maximum slope.

\end{document}